\pdfoutput=1 
\RequirePackage{ifpdf}
\documentclass{JINST}

\title{Development of a proton Computed Tomography Detector System}

\author{Md. Naimuddin$^a$\thanks{nayeem@cern.ch}, G. Coutrakon$^b$, G. Blazey$^b$, S. Boi$^b$, A. Dyshkant$^b$, B. Erdelyi$^b$, D. Hedin$^b$, E. Johnson$^b$, J. Krider$^b$, V. Rukalin$^b$, S. A. Uzunyan$^b$, V. Zutshi$^b$, R. Fordt$^c$, G. Sellberg$^c$, J. E. Rauch$^c$, M. Roman$^c$, P. Rubinov$^c$, P. Wilson$^c$\\
\llap{$^a$} Department of Physics \& Astrophysics, University of Delhi, Delhi 110 007, India\\
\llap{$^b$}Department of Physics, Northern Illinois University, DeKalb, IL 60115, USA\\
\llap{$^c$} Fermi National Accelerator Laboratory, Batavia, IL 60510, USA\\
}

\abstract{Computer tomography is one of the most promising new methods to image abnormal tissues inside the human body. Tomography is also used to position the patient accurately before radiation therapy. Hadron therapy for treating cancer has become one of the most advantageous and safe options. In order to fully utilize the advantages of hadron therapy, there is a necessity of performing radiography with hadrons as well. In this paper we present the development of a proton computed tomography system. Our second-generation proton tomography system consists of two upstream and two downstream trackers made up of fibers as active material and a range detector consisting of plastic scintillators. We present details of the detector system, readout electronics, and data acquisition system as well as the commissioning of the entire system. We also present preliminary results from the test beam of the range detector.}


\begin{document}

\section{Introduction}\label{sec:intro}
\begin{figure}[htp]
\begin{center}
\includegraphics[width=3.0in]{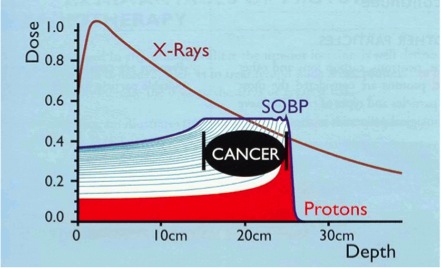}
\end{center}
\caption{\label{fig:Bragg_Peak} The schematic of the energy loss of protons in matter.}
\end{figure}

Hadron therapy for treating tumors is becoming increasingly popular. Hadrons beams reduce damage to healthy tissues by reducing the overall dose to the patient as compared to conventional x-ray therapy. The interaction of hadrons with matter is characterized by the formation of a Bragg peak where most of the energy is depositedand very little energy beyond the peak (as shown in Fig.~\ref{fig:Bragg_Peak}). Using this characterstic one can adjust the range and the energy of the beam to form the spread out Bragg peak (SOBP) right at the spot where the tumor is, and thus reducing the dose the healthy tissue. 

In order to fully utilize the properties of proton beams in radiation therapy, the proton stopping power within the human body must be known to a good accuracy prior to the treatment. This information is presently accessed through x-ray computed tomography (CT) of the patients. To convert the x-ray CT Hounsefield units to proton relative stopping powers, which are required by the treatment planning software, an empirically derived calibration function is used~\cite{Hounsfield}, which is specific to each x-ray CT machine. But, this calibration between the Hounsfield units and relative stopping powers is not unique because of the different dependence on Z and $Z/A$ ratio of x-ray and proton energy loss. This leads to an uncertainty~\cite{Schneider} of about 3 to 4\% in the location of the tumour at the time of the treatment. This uncertainty could be reduced by a direct measurement of the stopping power. This can be achieved if the patient is radiographed directly with the proton beam. The aim of the proton CT is to reduce this uncertainty to 1\% of the full range. 

X-ray CT also deposits considerable dose to healthy tissue during the scanning as can be seen Fig.~\ref{fig:Bragg_Peak}. With proton CT a three to five times reduction in dose compared to X-ray CT is possible. In addition, artifcats arising from high density dental or other implants can be minimized leading to higher quality images. In order to achieve these goals, a new proton CT scanner is being built and tested to acquire human head size images with scan times below 10 minutes. This proton CT scanner is being built at Northern Illinois University in collaboration with Fermi National Accelerator Laboratory in Batavia, IL and University of Delhi, Delhi, India.

\section{Proton Tomography Detector System}\label{sec:pctdet}
\begin{figure}[htp]
\begin{center}
\includegraphics[width=5.0in]{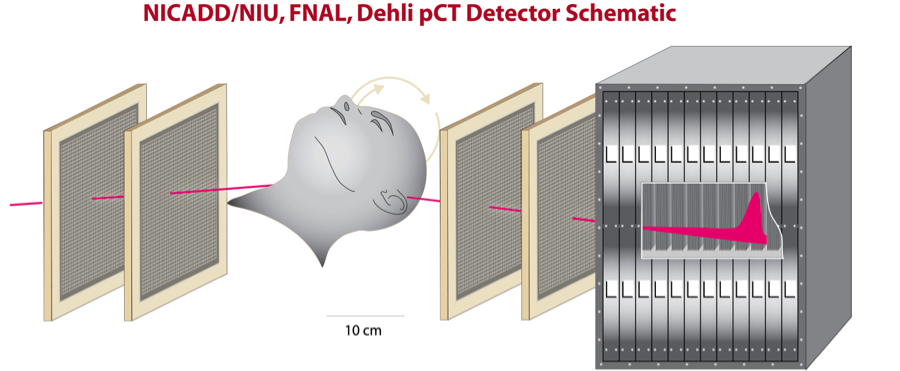}
\end{center}
\caption{\label{fig:pCT_Sketch} The sketch of second generation proton computed tomography system.}
\end{figure}

The Proton Computed Tomography (pCT) detector consists of two upstream tracking detectors, two downstram tracking detector and a range detector to measure the residual energy of the protons. The model of the pCT detector system is shown in Fig.~\ref{fig:pCT_Sketch}. The tracking detectors provides the histories of each proton in terms of its x, y and z coordinates for each hit in the detector. 

\begin{figure}[htp]
\begin{center}
\includegraphics[width=5.0in]{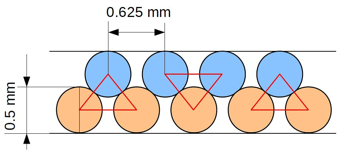}
\end{center}
\caption{\label{fig:Fiber_Layer} The bundeling of the Fibers for the tracking detector.}
\end{figure}

Each tracking detector is made up of 4 layers of 0.5 mm diameter polystyrene scintillating fibers by Kuraray~\cite{Kuraray} in the x-y plane. Fibers were cut into 20 and 30 cm length, then laid flat on a low density, 0.03 $g/cm^3$, 2 mm thick rohacell substrate with machine grooves. The fibers were glued to hold on to the rohacell in grooved spaces. Another layer was put on top in the valleys caused by the two adjacent fibers in the first layer, as shown in Fig.~\ref{fig:Fiber_Layer}, to avoid the gaps in detecting passing protons. Another two similar layers were put on the other side of the rohacell, but rotated by $90^{O}$. The first two planes are lying in -x to +x from end-to-end while the next two layers are lying in the -y to +y direction end-to-end. The entire assembly is supported on Techtron (carbon fiber) frames. 

For the readout, these fibers were bundled into triplets, as shown in Fig.~\ref{fig:Fiber_Layer}, which provides a pitch of 0.94 mm between the bundles. Each bundle is read out through Silicon photo-multipliers (SiPMs), manufactured by CPTA~\cite{SiPM}, which are mounted on blocks connecting each of them to a fiber triplet. One end of each fiber is coated with a reflective material while the other end is polished and mechanically pressed to an SiPM on a block. The root mean square (rms) spatial resolution of each tracker plane is given by the pitch divided by $\sqrt{12}$, or 0.27 mm. The integrated water equivalent thickness (WET) of each tracker is less than 1 mm. 

The range detector (or the calorimeter) is used for measuring the residual energy of the protons emerging from the phantom. It consists of 3.2 mm plastic scintillator plates as the active detector. We choose plastic scintillator as active material because of its low cost and excellent timing resolution. These detectors work by absorbing incident protons and then re-emitting scintillating photons of much longer wavelength which can then be made to fall on the photocathode tubes of the photomultipliers. 

The range detector is made up of 12 frames separated by 9.25 mm. Each frame consists of 8 layers of 3.2 mm thick (330 $mg/cm^2$) plastic scintillator made of polyvinyltolulene. There is one layer of aluminized mylar of thickness 6.25 $\mu$m (0.408 $mg/cm^2$) between the plates, before the plates and after the plates. The thickness of the aluminium coating on the mylar is approximately 2200 angstroms. A total of 9 layers of aluminized mylar sheets are used for each frame.


\section{Electronics \& Data Acquisition System}\label{sec:daq}
\begin{figure}[htp]
\begin{center}
\includegraphics[width=5.0in]{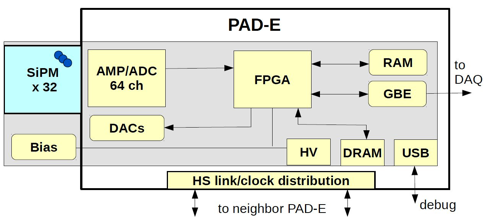}
\end{center}
\caption{\label{fig:pCT_Electronics} The read out electronics for the pCT.}
\end{figure}

The readout electronics consists of a custom board with preamps, digitizers, FPGAs, and ethernet readout (PAD-E) as shown in Fig.~\ref{fig:pCT_Electronics}. These boards can provide readout for up to 32 channels of SiPM in a $220~mm \times 100~mm$ format. The Same board is used for the readout of the trackers and the range calorimeter. The signals from the SiPMs are digitized after appropriate amplification and shaping at the rate of 12 bits per channel at 75 MSPS. The PAD-E is fully self contained and produces the bias for the SiPMs. The boards also contain the FPGA for processing all the data from the SiPMs,  provides memory for buffering up to 128 MB of data, and a gigabit ethernet interface for sending the data directly to the DAQ. The PAD-E is powered by a single 5V power supply and consumes power up to 15 W for 32 SiPM channels. The scanner is self triggered in the sense that any channel with signal above threshold will be time stamped and stored in a local buffer for the readout. A synchronous signal allows all the boards to provide a timestamp that is used by the DAQ to combine data from various parts of the detector for a single proton history. A synchronization signal which run across all the boards once per millisecond intiates a packet of data readout from PAD-E memory to DAQ memory through 1 Gbit/s ethernet with only little dead time penalty. The organization of the data into a millisecond time packets allows for a relatively small timestamp of 16 bits of 75 MHz clock cycle and also monitors the integrity of the data.

\begin{figure}[htp]
\begin{center}
\includegraphics[width=5.0in]{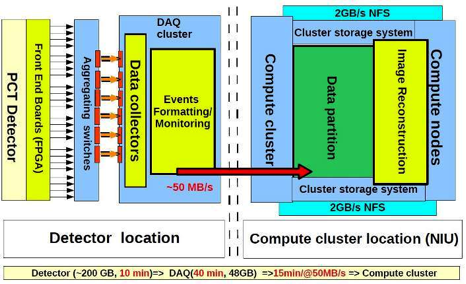}
\end{center}
\caption{\label{fig:pCT_DAQ} The data acquisition system for the proton computed tomography detector.}
\end{figure}

The data from the front end electronics are sent to the DAQ system, as shown in Fig.~\ref{fig:pCT_DAQ}, over a 1 Gbit/s ethernet lines using UDP protocol. Each proton histroy contains the data for all the eight tracker planes and the 96 scintillator tiles from the stack. It is estimated that each proton history generates about 25 bytes of data from the 8 hits in the eight tracking detectors and about 75 bytes from the 96 scintillator plates. The scan is done with the rotation of the phantom in steps of three degrees. For a 10 minute scan with 120 projection angles at a event rate of 2 million protons per second, it is expected that 200 MB/s data will be written to RAM by 24 data collectors running on six interconnected Linux workstations. After the scan the back end DAQ will write the data to disk for the further processing. The data is finally processed for the image reconstruction in the format such that it contains 4 X and 4 Y coordinates from the tracking detectors, Water Equivalent Path Length (WEPL)\cite{Hurley}, and the phantom rotation angle. 

\section{Summary}\label{sec:summary}
\begin{figure}[htp]
\begin{center}
\includegraphics[width=5.0in]{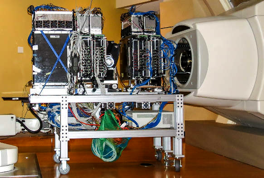}
\end{center}
\caption{\label{fig:pCT_Assembled} The full pCT detector at the test beam.}
\end{figure}

\begin{figure}[htp]
\begin{center}
\includegraphics[width=4.0in]{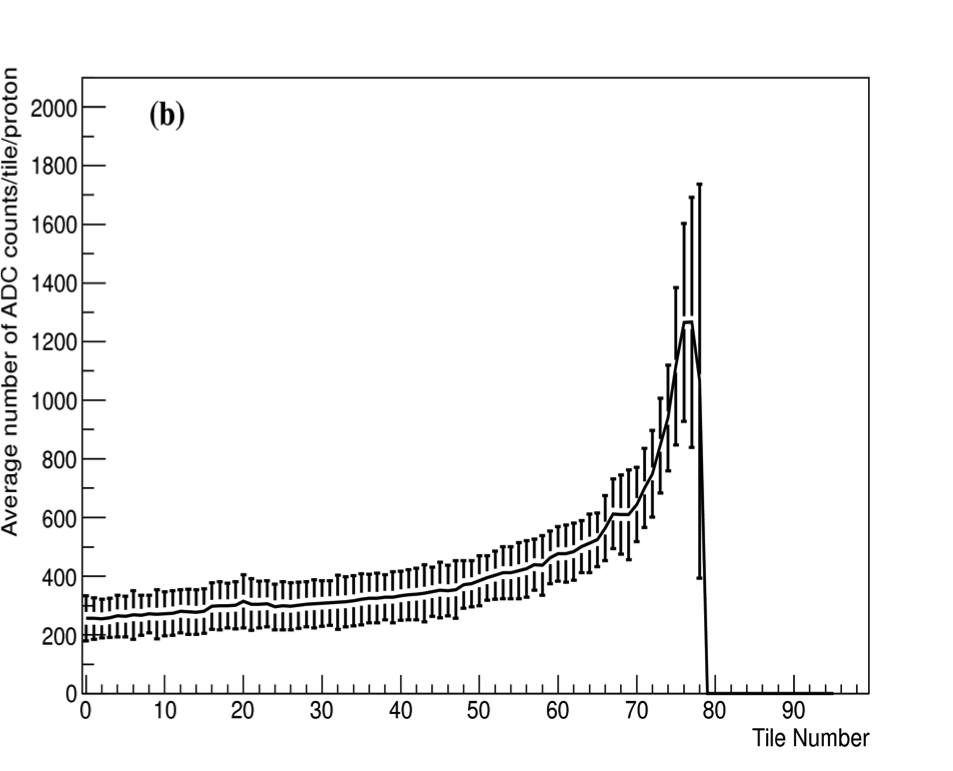}
\end{center}
\caption{\label{fig:pCT_BraggPeak} The energy profile in the range detector. The bragg peak can be clearly seen.}
\end{figure}

The proton CT scanner has been fully constructed and the entire detector system is being commissioned at the Central DuPage Hospital (CDH) in Warrenville, IL, USA with a 200 MeV proton beam. Fig.~\ref{fig:pCT_Assembled} shows the fully assembled pCT detector system at the CDH treatment room. The range detector was earlier commissioned in the same beam line. Fig.~\ref{fig:pCT_BraggPeak} shows an enery deposition profile as a function of the tile number. The Bragg peak could be clearly seen near tile number 78 for a 200 MeV proton beam. Afterwards, the full detector has been placed in the beamline. We have already collected data from several runs with varying energy (range). We are currently in the final stages of the data analysis. The image reconstruction algorithms are already in place, and is currently being tuned and tested on the MC data generated using the GEANT4 model of pCT. We hope to finalize the data analysis to obtain the radiographs and reconstructed images by early 2016.

\acknowledgments

We wish to thank the Loma Linda University Medical Center and Central DuPage Hospital for allowing us to use their facilities. We wish to thank the US Army Medical Research Activity Center in Ft. Detrick, MD for providing the funding to make this project possible. We also thank the University Grants Commission (UGC) and University of Delhi for providing the $R\&D$ grants at the Delhi University for carrying out part of these studies.

\end{document}